\documentstyle[11pt,twoside]{article}

\begin{document}

\newcommand{\dd}{\,{\rm d}}
\newcommand{\ie}{{\it i.e.},\,}
\newcommand{\etal}{{\it et al.\ }}
\newcommand{\eg}{{\it e.g.},\,}
\newcommand{\cf}{{\it cf.\ }}
\newcommand{\vs}{{\it vs.\ }}
\newcommand{\zdot}{\makebox[0pt][l]{.}}
\newcommand{\up}[1]{\ifmmode^{\rm #1}\else$^{\rm #1}$\fi}
\newcommand{\dn}[1]{\ifmmode_{\rm #1}\else$_{\rm #1}$\fi}
\newcommand{\upd}{\up{d}}
\newcommand{\uph}{\up{h}}
\newcommand{\upm}{\up{m}}
\newcommand{\ups}{\up{s}}
\newcommand{\arcd}{\ifmmode^{\circ}\else$^{\circ}$\fi}
\newcommand{\arcm}{\ifmmode{'}\else$'$\fi}
\newcommand{\arcs}{\ifmmode{''}\else$''$\fi}
\newcommand{\MS}{{\rm M}\ifmmode_{\odot}\else$_{\odot}$\fi}
\newcommand{\RS}{{\rm R}\ifmmode_{\odot}\else$_{\odot}$\fi}
\newcommand{\LS}{{\rm L}\ifmmode_{\odot}\else$_{\odot}$\fi}

\newcommand{\Abstract}[1]{{\footnotesize\begin{center}ABSTRACT\end{center}
\vspace{1mm}\par#1\par}}

\newcommand{\TabCap}[2]{\begin{center}\parbox[t]{#1}{\begin{center}
  \small {\spaceskip 2pt plus 1pt minus 1pt T a b l e}
  \refstepcounter{table}\thetable \\[2mm]
  \footnotesize #2 \end{center}}\end{center}}

\newcommand{\TableSep}[2]{\begin{table}[p]\vspace{#1}
\TabCap{#2}\end{table}}

\newcommand{\TableFont}{\footnotesize}
\newcommand{\TableFontIt}{\ttit}
\newcommand{\SetTableFont}[1]{\renewcommand{\TableFont}{#1}}

\newcommand{\MakeTable}[4]{\begin{table}[htb]\TabCap{#2}{#3}
  \begin{center} \TableFont \begin{tabular}{#1} #4 
  \end{tabular}\end{center}\end{table}}

\newcommand{\MakeTableSep}[4]{\begin{table}[p]\TabCap{#2}{#3}
  \begin{center} \TableFont \begin{tabular}{#1} #4 
  \end{tabular}\end{center}\end{table}}

\newenvironment{references}%
{
\footnotesize \frenchspacing
\renewcommand{\thesection}{}
\renewcommand{\in}{{\rm in }}
\renewcommand{\AA}{Astron.\ Astrophys.}
\newcommand{\AAS}{Astron.~Astrophys.~Suppl.~Ser.}
\newcommand{\ApJ}{Astrophys.\ J.}
\newcommand{\ApJS}{Astrophys.\ J.~Suppl.~Ser.}
\newcommand{\ApJL}{Astrophys.\ J.~Letters}
\newcommand{\AJ}{Astron.\ J.}
\newcommand{\IBVS}{IBVS}
\newcommand{\PASP}{P.A.S.P.}
\newcommand{\Acta}{Acta Astron.}
\newcommand{\MNRAS}{MNRAS}
\renewcommand{\and}{{\rm and }}
\section{{\rm REFERENCES}}
\sloppy \hyphenpenalty10000
\begin{list}{}{\leftmargin1cm\listparindent-1cm
\itemindent\listparindent\parsep0pt\itemsep0pt}}%
{\end{list}\vspace{2mm}}

\def\TYLDA{~}
\newlength{\DW}
\settowidth{\DW}{0}
\newcommand{\dw}{\hspace{\DW}}

\newcommand{\refitem}[5]{\item[]{#1} #2%
\def\REFARG{#3}\ifx\REFARG\TYLDA\else, {\it#3}\fi
\def\REFARG{#4}\ifx\REFARG\TYLDA\else, {\bf#4}\fi
\def\REFARG{#5}\ifx\REFARG\TYLDA\else, {#5}\fi.}

\newcommand{\Section}[1]{\section{#1}}
\newcommand{\Subsection}[1]{\subsection{#1}}
\newcommand{\Acknow}[1]{\par\vspace{5mm}{\bf Acknowledgements.} #1}
\pagestyle{myheadings}

\newfont{\bb}{timesbi at 12pt}

\def\thefootnote{\fnsymbol{footnote}}

\begin{center}
{\Large\bf Relativistic Effects in Proper Motions of Stars \\
\vskip3pt
Surrounding the Galactic Center}

\vskip1cm
{\bf M. Jaroszy{\'n}ski}
\vskip5mm
{Warsaw University Observatory, Al.~Ujazdowskie~4, 00-478~Warszawa, Poland\\
e-mail: mj@sirius.astrouw.edu.pl }
\vskip5mm
Received November 16, 1998
\end{center}
\vskip 10mm

\Abstract{We simulate the astrometric observations of stars moving close to
the black hole in the Galactic Center. We show, that for orbits $\le
10^3~{\rm AU}$ and position measurements with the accuracy of the Keck
Interferometer, the periastron motion of elliptical orbits will be
measurable. The models of star trajectories neglecting the periastron
motion will be easy to reject with the high confidence level. The
measurement of orbital elements and the periastron motion can be
effectively used as an independent estimate the distance to the Galactic
Center. The effects of orbit precession may be visible in some cases. 
The effects of gravitational radiation are completely negligible as well as
the influence of the black hole rotation on the propagation of light.
}
{galaxies:black holes - galaxies: individual (Milky Way) - gravitational
lensing - relativity} 


\Section{Introduction}

The proper motion studies of stars in the Galactic Center (Eckart et al.
1995, Genzel et al. 1996, 1997, Ghez et al. 1998) have shown the astonishing
accuracy of astrometric observations in the near infrared $K$ band. 
Their studies have 
proved the existence of a very compact dark mass, most likely a black hole
of  $\sim 2.6 \times 10^6~{\rm M}_\odot$. The closest investigated star is
at the projected distance $100~{\rm mas}$, which corresponds to  
$\sim 850~{\rm AU}$. The gravitational radius for the quoted mass, 
$GM/c^2 = 0.025~{\rm AU}$ and corresponds to $3~{\rm \mu as}$. Thus the
motion at $\sim 3 \times 10^4$ gravitational radii from the mass is already
observed. 

Jaroszy{\'n}ski \& Paczy{\'n}ski (1998, hereafter JP98) have investigated
the possibility of finding orbital parameters for stars near the Galactic
Center. According to their work the systematic observations of the closest
star included in the proper motion studies can be sufficient to define its
orbit in $\sim 10~{\rm y}$ with the accuracy allowing for the determination
of the central mass with an error smaller then the present estimates.
Increased accuracy of astrometric measurements and ability to observe
fainter stars which may be found closer to the center, will allow for still
more accurate determinations in shorter time. 

The speckle interferometry with
the Keck telescope (Ghez et al. 1998) reaches the resolution 
$\sim 50~{\rm mas}$ and the accuracy of $\sim 2~{\rm mas}$ in position
measurements.  The Keck Interferometer, which is now under construction
(van Belle \& Vasisht 1998) will have $\sim 5~{\rm mas}$ resolution and 
$\sim 20~{\rm \mu as}$ ($0.17~{\rm AU}$) astrometric accuracy. 
It will be able to measure the 
position of a faint point object with $K \le 22^m$ if a bright $K \le 14^m$
star can be found within a circle of the radius $\le 20~{\rm arcsec}$
around it. 
Such measurement accuracy, up to few gravitational radii,  suggests the
possibility of investigating relativistic corrections to the motion of
stars sufficiently close to the central mass. We include the periastron
motion, the precession of the orbit, the gravitational radiation from the
star-central mass binary, and the bending of rays in our study.

In this paper we assume that there is a $\sim 2.6 \times 10^6~{\rm M_\odot}$ 
black hole in the Galactic Center and we investigate the measureability of
the relativistic effects in its vicinity. 
We consider orbits of stars of sizes $50$ to $1000~{\rm AU}$, which
correspond to periods $0.219$ to $19.6~{\rm y}$.
Following JP98 we use Monte Carlo
method to simulate the observations of the stars. We use minimization
algorithms to fit the model trajectories to the synthetic data sets.
Using models with different level of sophistication and comparing the results
we are able to find which relativistic effects must be included in the
interpretation of observations and which are below the current and near
future detection limits.

In the next Section we describe the motion of particles and photons in the
weak gravitational field of a rotating body. In Section~3 we present the
methods of simulating the observations and procedures of parameter fitting.
We also present the main results of this paper - the range of orbit sizes for
which the study of relativistic effects is possible with given astrometric
accuracy. We also calculate the accuracy of the distance estimate based on
the measured periastron motion.
In Section~4 we estimate the chance that a star (stars) bright
enough to be followed by the Keck Interferometer may be found close enough
to the Galactic Center, so the relativistic effects are measurable. The
conclusions follow in the last Section.

%
\Section{Equations of motion in the weak field approximation}

We describe the gravitational field far from a rotating  black hole in the
isotropic coordinates $t$, $x$, $y$, and $z$. We use the geometrical units
for mass ($M$) and angular momentum ($J$) of the hole:
\begin{equation}
m \equiv {GM \over c^2}~~~~~~~j \equiv {J \over Mc}
\end{equation}
where $G$ is the gravity constant and $c$ is the velocity of light. 
The black hole angular momentum is directed along ${\bf e_z} \equiv 
{\bf k}$. The observer is in the $({\bf e_x},{\bf e_z})$ plane.
In the
approximation preserving terms of the order $\sim r^{-2}$, the lowest
including effects of the black hole angular momentum, the metric takes the
form (Landau \& Lifshitz 1973):
$$
ds^2=-\left(1-2{m \over r}+2{m^2 \over r^2}\right)dt^2
+4{jmy \over r^3}dtdx-4{jmx \over r^3}dtdy
$$
\begin{equation}
~~~~~~~~~~+\left(1+2{m \over r}+{3 \over 2}{m^2 \over r^2}\right)
\left(dx^2+dy^2+dz^2\right)
\end{equation}

\subsection{Orbits of Stars}

In the weak field approximation the orbit of the star can be described as
a classical ellipse of semimajor axis $a$ and eccentricity $e$ 
in a non-inertial frame of reference with coordinates
$(x_1,y_1,z_1)$. 
We assume that the black hole is at the origin of this
coordinate system, the orbit is at the $({\bf e_{x_1}},{\bf e_{y_1}})$ 
plane, and
${\bf e_{x_1}}$ points toward the orbit periastron. The orientation of the
frame is such that the orbital angular momentum is directed along the
positive direction of ${\bf e_{z_1}} \equiv {\bf n}$. The inclination $i$
of the orbital plane relative to the equatorial plane is given by the
condition $\cos i=({\bf k}{\bf n})$. The ascending node of the orbit is at
the position angle $\Omega$ measured from the $x$ axis. The periastron is
at the angle $\omega$ from the ascending node. Finally the position of the
star on the orbit, as measured by the eccentric anomaly $u$, is given by
the Kepler equation:
\begin{equation}
{2\pi \over P}(t-t_0)=u-e\sin u
~~~~~~x_1=a(\cos u-e)
~~~~~~y_1=a\sqrt{1-e^2}\sin u
\end{equation}
where $P$ is the orbital period and $t_0$ is the time of the passage
through the periastron. 

The Lense-Thirring effect causes the precession of the orbital plane and
the change of the periastron location. As a result the line of nodes
rotates with the angular velocity:
\begin{equation} 
{\dot \Omega}_{\rm prec}={2jmc \over a^3(1-e^2)^{3/2}}~~~~~~
{\dot \Omega}={\dot \Omega}_{\rm prec}\sin i
\end{equation}
and the periastron position changes with the rate:
\begin{equation}
{\dot \omega}_{\rm prec}=-3{\dot \Omega}_{\rm prec}\cos i
\end{equation}
Independently of the black hole angular momentum, the periastron of the
orbit advances with the angular velocity:
\begin{equation}
{\dot \omega}_{\rm per}={3m^{3/2}c \over a^{5/2}(1-e^2)}
\end{equation}
The resulting evolution of the orbit orientation is given as:
\begin{equation}
\Omega=\Omega_0+{\dot \Omega}(t-t_0)~~~~~~~~
\omega=\omega_0+({\dot \omega}_{\rm per}
                      -3{\dot \Omega}_{\rm prec}\cos i)(t-t_0)
\end{equation}
where $\Omega_0$ and $\omega_0$ are the initial values of the angles.
The star location at any reference frame can be calculated after a
straightforward coordinate transformation based on the equation:
\begin{equation}
{\bf r}(t)=x_1(t)~{\bf e_{x_1}}(t)+y_1(t)~{\bf e_{y_1}}(t)
\end{equation}
Suppose an observer is located in the $(x,z)$ plane at the angle $\Theta$
measured from the hole rotation axis. The observed position of the star on
the sky is given as:
\begin{equation}
Y(t_{\rm obs})=y(t)~~~~~~
Z(t_{\rm obs})=z(t)\sin\Theta-x(t)\cos\Theta
\end{equation}
where we have chosen coordinates in the sky with $Z$ axis along the
projection of the black hole angular momentum and $Y$ axis along the $y$
axis of our frame. The position angle $\psi$, which gives the orientation
of the $Z$ axis on the sky is another parameter of the problem. The time of
the observation $t_{\rm obs}$ depends on the 
source position and is given (up to an additive constant) as:
\begin{equation}
ct_{\rm obs}=ct-x(t)\sin\Theta-z(t)\cos\Theta
\end{equation}

\subsection{The Influence of Gravitational Lensing}

In the typical lensing situation the source and the observer are far from
the lens, but the rays pass through its close vicinity. Such case of
lensing near the Galactic Center has been considered by Wardle \&
Yusef-Zadech (1992), Jaroszy{\'n}ski (1998), and Alexander \& Sternberg
(1998). In such a case one can
describe a ray as two segments of a straight line deflected in the lens
plane. In our case the distance of the source from the lens is of the same
order as the encounter parameter for rays and the standard gravitational
lensing formalism can not be employed. Instead we consider the null
geodesics in the metric of Eq.~2.

In the zeroth order approximation a ray is a straight line. We assume that
the point of the closest approach to the hole is at ${\bf b_0}$. 
Suppose the ray is propagating
along the unit vector ${\bf l}$ and let $l$ be the length measured along
the ray from the point of the closest approach. 
Since we are using the weak
field approximation, we have to limit ourselves to the case $b_0 \gg m$. For
a unit energy photon propagating along the ray the components of the four
momentum are: $p_t=-1$, $p^t=1$, $p_l=1$, $p^l=1$, and all other vanish.

In the first order approximation we include the terms proportional to $m/r$
in the metric, which preserves its spherical symmetry. Due to this symmetry
the ray remains in the plane $({\bf b_0},{\bf l})$ and is deflected toward
the hole:
\begin{equation}
{d \over dl}p_b={1 \over 2}g_{tt,b}p^tp^t+{1 \over 2}g_{ll,b}p^lp^l
=-{2mb_0 \over r^3}
\end{equation}
where $r=\sqrt{b_0^2+l^2}$. Integrating the above equation with the boundary
condition at the observer position: $p_b=0$ for $l=\infty$, one gets
\begin{equation}
p_b(l)={2m \over b_0}\left(1-{l \over \sqrt{b_0^2+l^2}}\right)
\end{equation}
For sources far behind the lens the photon momentum changes its direction
by an angle 
$(p_b(+\infty)-p_b(-\infty))/p_l=-4m/b_0$, which is a standard result for
a ray deflection.
The photon momentum perpendicular to the ray is a first order quantity, so
$p^b=p_b$. Integrating again and assuming $b(+\infty)=b$ we have:
\begin{equation}
{d \over dl}b(l)=p^b~~~~~~
b(l)=b-{2m \over b_0}\left(\sqrt{b_0^2+l^2}-l\right)
\end{equation}
where the second relation is the analog of the lens equation.
For $l=0$ one has $b_0=b-2m$ which is the relation between the observed
source position ($b$) and the encounter parameter ($b_0$). In general the
source is at the distance $b_s$ from the optical axis, at the position $l_s$
along the ray. The value of the encounter parameter $b_0$ can be obtained
as a solution to the equation $b(l_s)=b_s$. In a typical case, when 
$b_s \gg m$, the same is true of $b$ and $b_0$ and one has approximately:
\begin{equation}
b=b_s+{2m \over b_s}\left(\sqrt{b_s^2+l^2}-l\right)
\end{equation}
The formula is valid for $\vert l \vert \le b_s^2/m$. In the language of
gravitational lensing it means, that the source should be at a distance
much larger than Einstein radius from the optical axis.

The second order terms ($\sim m^2/r^2$) in the diagonal metric components
introduce only quantitative corrections to the lens equation. The off
diagonal terms introduce the dependence of geodesics on the black hole
angular momentum. The deflection of rays is not necessarily toward the hole.
Since the metric components of interest are of the second order, the
influence of the hole angular momentum can be calculated along the zeroth
order rays.  Only the deflection perpendicular to the line of sight is
interesting. In the observer's coordinates the geodesic equations are:
\begin{equation}
{d \over dl}p_Y={2jm\sin\Theta \over r^3}-{6jmY_0^2\sin\Theta \over r^5}
\end{equation}
\begin{equation}
{d \over dl}p_Z=-{6jmY_0Z_0\sin\Theta \over r^5}
\end{equation}
where the encounter vector is given as ${\bf b_0}=(Y_0,Z_0)$ and the
distance from the hole as $r=\sqrt{b_0^2+l^2}$. 
The integration gives:
\begin{equation}
\delta Y={2jm\sin\Theta \over b_0^4}(Z_0^2-Y_0^2)
\left(\sqrt{b_0^2+l^2}-l\right)
+{2jm\sin\Theta Y_0^2 \over b_0^2\sqrt{b_0^2+l^2}}
\end{equation}
\begin{equation}
\delta Z=-{4jmY_0Z_0\sin\Theta \over b_0^4}
\left(\sqrt{b_0^2+l^2}-l\right)
+{2jmY_0Z_0\sin\Theta  \over b_0^2\sqrt{b_0^2+l^2}}
\end{equation}
where $\delta Y$, $\delta Z$ denote the angular momentum induced shift of
the ray from a trajectory neglecting these effects. Even for the maximally
rotating black hole ($j=m$), the shifts are of the order 
$\sim m~m/b_0 \ll m$ and we neglect them in further calculations.

\subsection{The Gravitational Radiation}

The gravitational radiation lowers the energy of a binary system of masses
leading to the orbit narrowing and shortening of the orbital period. For
two masses $M_1$ and $M_2$ at a distance $r$ from each other, moving on
circular orbits, one has (Landau \& Lifshitz 1973): 
\begin{equation}
{\dot r} = -{64G^3M_1M_2(M_1+M_2) \over 5 c^5r^3}
\end{equation}
We estimate the relative change of the orbit size for a star with mass 
$M_2=2.6 M_\odot$ revolving around a black hole 
of the mass $M_1=2.6 \times 10^6~M_\odot$. 
We get
\begin{equation}
{\vert{\dot r}P\vert \over r}
\approx 10^{-13}~\left({100 {\rm AU} \over r}\right)^{2.5}
\end{equation}
which means that the gravitational radiation can be neglected for a wide
range of star masses and orbit sizes.

\section{Observability of the Relativistic Effects}

We simulate the astrometric observations of stars on elliptic orbits around
the Galaxy Center black hole. We check for which range of orbital periods
the measurement of the star proper motion with given positional accuracy is
sufficient to measure the rate of the periastron motion. We also check,
whether the precession of the orbital plane, which may be present if the
black hole is rotating, can cause any measurable effects.

The simulations consist of two parts. First we choose the
physical parameters of the orbits ($a$, $e$, $P$), their orientation in
space ($i$, $\Omega_0$, $\omega_0$) and time ($t_0$). The value of the
semimajor axis is the main parameter of this study and it covers a range of
values. We use the linear measure of the orbit semimajor axis $a$, while
the quantity more directly related to observations is the corresponding
angle $\alpha$ ($a \equiv \alpha R$, where $R$ is the distance to the
Galactic Centre, for which we use $R=8.5~{\rm kpc}$).
The orbital period is related to the orbit size, since we assume
that the black hole has the mass $M=2.6 \times 10^6~M_\odot$ (Ghez
et al. 1998). Other parameters are chosen at random for each orbit. We
assume that any orientation of the orbit in space and any initial phase of
the orbital motion have equal probabilities. The initial time of the
measurement seems to be unimportant, but due to the fact that the
conditions for astronomical ground observations of Sgr A$^*$ do change
through the year, we treat $t_0$ as another random variable. For simplicity
we assume, that the observations are possible from March through September
 or from the 10-th to the 40-th week of the year. We do not introduce any
seasonal or random dependence of the accuracy of the astrometric
measurements on time. The rate of the periastron motion ${\dot \omega}$
and the  precession angular velocity ${\dot \Omega}$ can be calculated,
when other orbital parameters are known.

The equations of Sec.~2 can be used to obtain the ``true'' trajectory of the
star projected into the plane of the sky 
${\bf X}
(t;t_0,a,e,p,i,\Omega_0,\omega_0,{\dot \Omega},{\dot \omega},\Theta,\psi)$,  
or ${\bf X}(t)$ in short notation.
Observations of the star in the instants of time $\{ t_j\}$ give the
measured values of its positions in the sky $\{ {\bf X_j} \}$. We simulate
the process of observation assuming that
\begin{equation}
{\bf X_j}={\bf X}(t_j)+\delta{\bf X_j}
\end{equation}
where $\delta{\bf X_j}$ are the errors introduced by observations. We
assume that each component of the position vector is measured with 
errors which are normally distributed with dispersion $\sigma$ and
vanishing mean value. Equivalently the distribution of $\delta{\bf X_j}$ is
given as:
\begin{equation}
P\{ \vert \delta{\bf X_j}\vert > s\}=\exp\left(-{s^2 \over 2\sigma^2}\right)
\end{equation}
and the direction of $\delta{\bf X_j}$ has a uniform distribution. Using
Monte Carlo method we obtain synthetic data sets $\{ {\bf X_j}\}$
representing the sequences of astrometric observations with noise.

The next step is to fit a model to each synthetic data set. We fit
orbital parameters to the observations by minimizing the expression:
\begin{equation}
\chi^2=\sum_{j=1}^{N}~
{({\bf X_j}-{\bf X}(t_j;
t_0,a,e,P,i,\Omega_0,\omega_0,{\dot \Omega},{\dot \omega},\Theta,\psi))^2
\over \sigma^2}
\end{equation}
The parameters fitted to a synthetic data set are different from the
original ``true'' parameters of the orbit. Many simulations of synthetic
data sets for the same orbit give the scatter in estimated parameters.
This ``bootstrap'' method (Press et al. 1988) is a practical way to
estimate the accuracy of parameter fitting in the case of real
observations.

The synthetic data sets we obtain using Monte Carlo method are always based
on calculations including the effects of the periastron
motion, the precession of the orbit and the bending of rays by
gravitational lensing. To check whether these effects  are observable we
compare the quality of the fits obtained with models including or
neglecting them. 
We start from the
simplest model which neglects both the precession and periastron motion of
the orbit. Since we are limiting ourselves to the first order effects in
gravitational lensing, the angular momentum of the black hole has no
influence on the visual orbit and the reference frame defined by the hole
rotation axis looses its observational basis. The inclination of the orbit
should now be defined relative to the plane of the sky. The position of the
line of nodes can be measured from the declination circle.
We obtain the fits by minimizing $\chi^2$ again, but using a simplified
model ${\bf {\hat X}}(t;t_0,a,e,P,i,\Omega_0,\omega_0)$ instead of the full
model depending on the higher number of parameters. This can also be
achieved by fixing the values of some parameters (${\dot \Omega} \equiv 0$,
${\dot \omega} \equiv 0$, $\Theta \equiv 0$, and $\psi \equiv 0$) in the
full model. 
As we show in Fig.~1 the model neglecting the periastron  motion and the
precession of the orbit can be rejected for sufficient accuracy of position
measurements and short orbits.

The precession of the orbit is a weaker effect than the periastron
motion. If the observations span only a few orbital periods, which is the
case of our simulations, the two effects can be difficult to
distinguish. To clarify this point we try a model which 
includes periastron motion but neglects the precession, 
${\bf {\hat X}}(t;t_0,a,e,P,i,\Omega_0,\omega_0,{\dot \omega})$.
Our calculations show, that in this case the successful fits are possible.
By successful we mean the fits with sufficiently low value of $\chi^2$:
\begin{equation}
\chi^2 \le \chi^2_{0.95}(m)
\end{equation}
where $m=2N-8$ is the number of the degrees of freedom for $N$ observed
positions and $8$ parameters fitted. The subscript denotes the confidence
level. (See also the dotted line on Fig.~1.)
 
Using the fitted orbital parameters one can calculate the mass of the
central body with the help of the Third Kepler Law:
\begin{equation}
m={4\pi^2a^3 \over c^2P^2}
\end{equation}
 Assuming that the
central body is not rotating, one has the expected periastron motion per one
revolution:
\begin{equation}
\Delta \omega_{\rm exp}={\dot \omega}_{\rm per}P={6\pi m \over a(1-e^2)}
\end{equation}
We introduce another variables, independent of orbit eccentricity, which
also measure the periastron motion:
\begin{equation}
Q_{\rm exp}=\Delta \omega_{\rm exp}(1-e^2)~~~~~~
Q_{\rm fit}={\dot \omega}_{\rm fit}P_{\rm fit}(1-e^2)
\end{equation}
The first variable is based on the theory and fitted values of mass and
semimajor axis of the orbit. The second is based on the fitted values of
the periastron motion and the orbital period. For the nonrotating
central body the variables should be equal. If the central body rotates,
they should be different. Since both variables are based on the fitting
procedure and "observations" with errors we can only compare their
averaged values. We define:
\begin{equation}
D={\langle \vert Q_{\rm fit}-Q_{\rm exp}\vert \rangle 
\over \langle Q_{\rm exp} \rangle }
\end{equation}
\begin{equation}
S=
{\sqrt{\langle \left(Q_{\rm exp}-\langle Q_{\rm exp}\rangle\right)^2\rangle}
\over \langle Q_{\rm exp} \rangle }
\end{equation}
where the averages are taken for orbits of the same ``true'' semimajor axis
$a_0$ and the same accuracy of astrometric observations $\sigma$.
In Fig.~2 we compare the fitted and expected rates of periastron motion. We
consider the case of stars moving in the gravitational field of a
nonrotating black hole (left panel) and the case of maximally rotating hole
(right panel). The left panel shows that the difference between the fitted
and expected values scales linearly with the error in astrometric
measurements. The same is true of the dispersion in the expected rate of
periastron motion. This shows, that the differences are statistical in
nature. The plots also show that the ``observations'' which cover a given
number of rotational periods (5-10 in most cases) allow for a more accurate
fit to the  orbit semimajor axis and period than for a fit to the rate of the
periastron motion. The right panel shows a nonlinear behavior. Comparing 
Eqs.~4, 5, and 6 one can get a rough estimate:
\begin{equation}
{\Delta \omega_{\rm prec} \over \Delta \omega_{\rm per}} 
\sim \sqrt{j^2 \over ma} = \sqrt{m \over a} 
\end{equation}
where we have assumed $\vert\cos i \vert=0.5$ and neglected the factor
dependent on the eccentricity. The second equality is valid for the maximally
rotating hole ($j=m$). The presence of the precession in the ``true''
motion, which is not accounted for in the model causes a systematic
difference between the rates of the periastron motion estimated in two
ways. This difference remains finite when the error in measurements becomes
very small. Its value is in agreement with the above formula for the orbits
we consider. (Orbits with $a=50$, $150$, and $250~{\rm AU}$ 
or $P=0.22$, $1.14$, and $2.45~{\rm y}$ are considered here.) 

The expected contribution to the periastron motion from the precession 
is smaller than $\sim 1$~\% for sufficiently wide orbits 
($a \ge 250~{\rm AU}$), even for the maximally rotating central black
hole. If we neglect this effect, Eqs.~25,~26 can be used as two independent
methods of estimating the mass. After substitution we get the relation
between the fitted variables of our model:
\begin{equation}
\Delta \omega ={24\pi^3 \over c^2}~{a^2 \over (1-e^2)P^2}
\end{equation}
In our approach we generate the synthetic data sets using a fixed distance
to the Galactic Center, which makes the linear ($a$) and angular ($\alpha$)
measures of the semimajor axis indistinguishable. In reality the angular
size of the orbit $\alpha$ is fitted directly, while its physical dimension
can be calculated for the known distance to the object. Substituting
$a=\alpha R$ to the above equation we get:
\begin{equation}
R=\sqrt{(1-e^2)\Delta \omega \over 24\pi^3}~{cP \over \alpha}
\end{equation}
where all the variables in the RHS can be obtained from a fit to the
observations. Thus we have a method of estimating the distance to the
Galactic Center independent of any ``standard candles''.

We have estimated the error in the distance found by the above method for a
limited number of simulations. We have assumed that the black hole is
maximally rotating, so the errors resulting from the precession (which is
unaccounted for in the fitting procedure) are contained in our analysis.
The most important contribution to the
error comes from the measurement of the periastron motion despite the square
root dependence of the distance on this variable. The distances calculated
for orbits of given size are scattered. We find the median value of fitted
distances $R_{\rm med}$ and such $\Delta R$, that 68\% of the results
belongs to $[R_{\rm med}-\Delta R,R_{\rm med}+\Delta R]$. The plots of the
relative error in the distance measurement are given in Fig.~3. Only the
position measurements with errors smaller than $\sim 100~{\rm \mu as}$ can
give the distance estimate with the accuracy better than $\sim 10{\rm \%}$.

\Section{Stars Very Close to the Galactic Center}

The presence of stars at distances of few hundreds astronomical units from
the Galactic Center is necessary to measure the effects we consider. A more
detailed consideration of this subject can be found in JP98. The
observations of Eckart \& Genzel (1997), Genzel et al. (1996, 1997) and
Ghez et al. (1998) show the presence of the dense central star cluster. The
subset of stars with measured proper motions, which is seen close to the
center in projection, is also close to it in 3D, since the proper
velocities of stars are related to the projected distance - a fact hard to
understand for background or foreground objects. These stars are relatively
bright ($K \le 17^m$), but 
their sample is not complete (Ghez et al. 1998). This makes the following
reasoning a bit risky. Assuming that the stars with measured proper motions
are otherwise typical we can postulate that their luminosity function is
the same as that of the central star cluster. The integral luminosity
function for the Galactic Center in $K$ can be obtained from papers of Blum
et al. (1996) and Holtzman et al. (1998). It has a shape 
$N(\ge L_K) \sim L_K^{-\beta}$, with $\beta=0.875$, and flattens at 
$K \approx 21^m$. Thus going from $K=17^m$ to $K=21^M$ makes the star volume
density $\sim 25$ times higher and the typical distances between the stars
in 3D become $\sim 3$ times smaller. Since the closest to the center
observed star is at the distance $\sim 850~{\rm AU}$, one can expect few
fainter stars to be even closer. Thus the presence of observable stars at
required distance from the center is likely.

\Section{Conclusions}

We have investigated the observability of relativistic effects in motion of
stars in the vicinity of the Galactic Center assuming the presence of
massive black hole there and the accuracy of
astrometric position measurements up to $\sim 20~{\rm \mu as}$.
We have shown that the gravitational radiation from the star-black hole
binary and the second order effects in the deflection of rays related to
the angular momentum of the black hole are completely negligible. The only
robust effect is the motion of the orbit periastron. Systematic
observations of a star orbit, covering $\sim 25~{\rm y}$ or $\sim 5$
orbital periods (whichever takes shorter) are sufficient to measure the
rate of the periastron motion. With the accuracy of position measurements
$\sim 0.1~{\rm mas}$ or better, it is possible to reject models neglecting
the periastron motion for orbits of semimajor axis up to 
$\sim 500~{\rm AU}$. With the highest accuracy expected for the Keck
Interferometer ($\sim 20~{\rm \mu as}$) it will be possible to measure the
periastron motion for orbits up to $a \sim 10^3~{\rm AU}$.

If the black hole in the Galactic Center is rotating it should cause the
precession of the orbits of stars. We investigate this effect for orbits of
the size $a=50~{\rm AU}$ to $1000~{\rm AU}$. The effect is weak. The models
of orbits neglecting the precession can not be rejected on the basis of the
$\chi^2$ value. This is probably due to the fact that precession of the
orbit, when observed through a small number of revolution periods is hard
to distinguish from the periastron motion. (Both effects change the
direction of the ellipse axes in space. Precession can also change the
visual shape of the orbit, since it changes the angle between the line of
sight and the orbital plane, but this is a very slow process.)
The effects of precession can be seen indirectly as a discrepancy between
the theoretical rate of periastron motion for a nonrotating black hole of
estimated mass and the rate actually measured. 

The measurements of the orbit elements (size, eccentricity, and period) and
of the rate of the periastron motion can be used to estimate the distance
to the source. This method is independent of any standard candles. The
systematic observations of a star moving at distances $\le 10^3~{\rm AU}$
from the Galactic Center with the maximal positional accuracy of the Keck
Interferometer would give its distance up to few percent.

\Acknow{I thank Bohdan Paczy{\'n}ski for many useful discussions and the
anonymous referee for his suggestions.   This work was supported in part by
the Polish State Committee for Scientific Research grant 2-P03D-012-12.
}


\section{Note Added in Proof}

The highest astrometric accuracy of the Keck Interferometer 
($\sim 20~{\rm \mu as}$) will in fact be limited to relatively bright
objects with $K \le 17.6$. Thus the measurement of the periastron motion
will probably be possible only for stars already discovered. The best
candidate seems to be the star S0-1 from the Ghez et al. (1998) catalog. 
For the fainter stars the accuracy of position measurements will be much
worse, $\sim 3~{\rm mas}$, not adequate for the following of the periastron
motion. The results of our calculations can  still  be applied to
observations done with other instruments providing high astrometric
accuracy for faint objects, which may become operational in the future. 
I am grateful to Dr. Gerard T. van Belle for pointing to me my wrong
interpretation of the Keck Interferometer technical data.


\begin{figure}[p]
\vspace{13.0cm}
\includegraphics{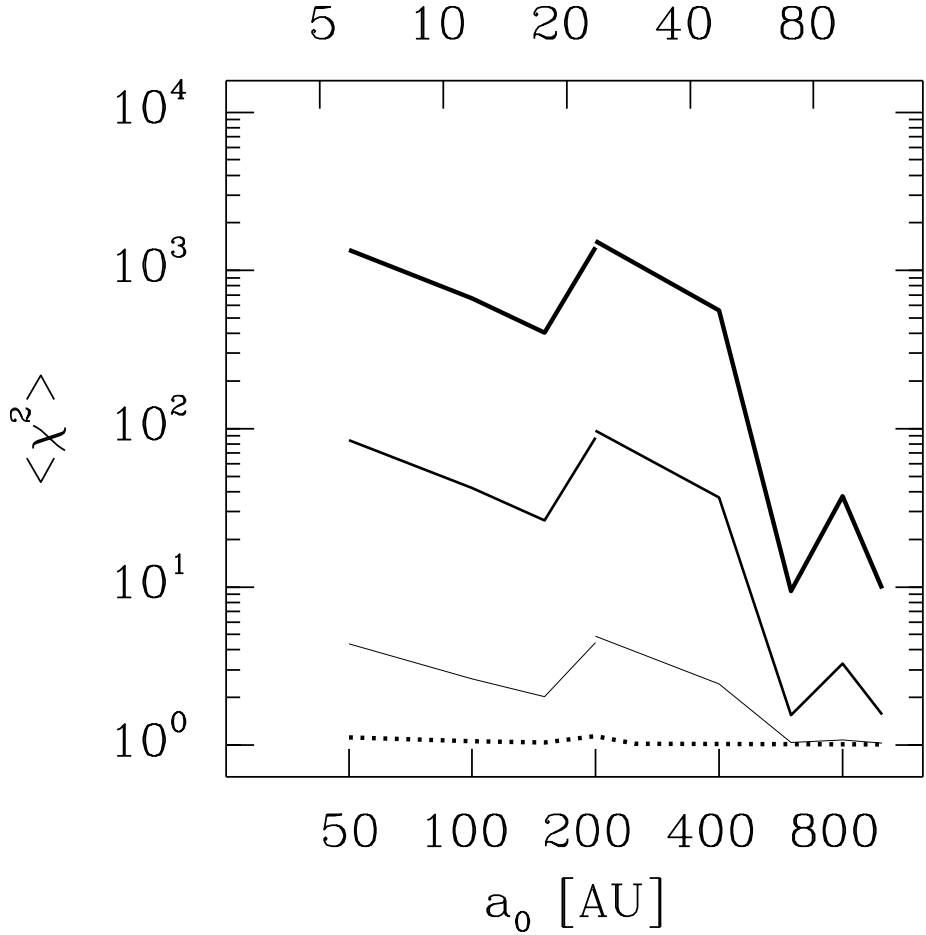}
\caption{The averaged $\chi^2$ per one degree of freedom for models
neglecting periastron motion (solid lines) compared with the averaged
$\chi^2$ for models taking into account the pariastron motion (dotted
line). The fits were attempted for orbits of
``true'' semimajor axis $a_0=50$ to $1000~{\rm AU}$, which corresponds to
periods $P_0=0.22$ to $19.6~{\rm y}$ and angular sizes $6$ to $118~{\rm
mas}$.  The results are shown
for simulated observations with the astrometric accuracy of
$\sigma=0.17~{\rm AU}$ ($20~{\rm \mu as}$, thick line), 
$0.68~{\rm AU}$ ($80~{\rm \mu as}$, medium line), and 
$3.4~{\rm AU}$ ($400~{\rm \mu as}$, thin line). 
Each model is based on 25 ``observations'' of
star position which are not more frequent than five per period and not less
frequent then once a year. That implies the dependence of the ``observation
strategy'' on the orbit size.
}
\end{figure}

\begin{figure}[p]
\vspace{10.0cm}
\includegraphics{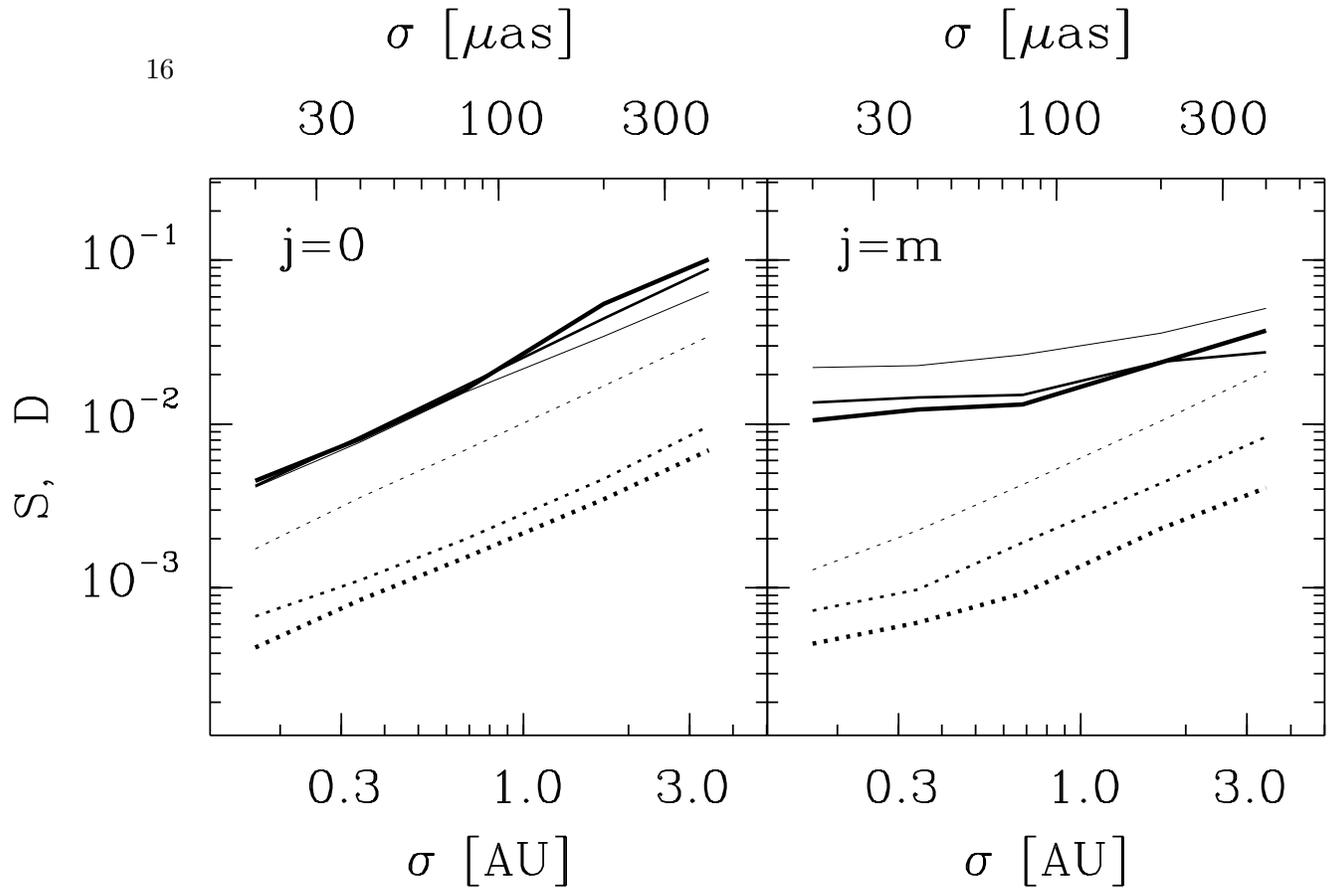}
\caption{The relative difference between the fitted rate of the periastron
motion and the rate expected for a nonrotating black hole 
of the same mass (solid lines) as a function of the accuracy of astrometric
measurements. Also shown is the dispersion of the latter
quantity (dotted lines). On the left panel the case of a nonrotating black
hole is shown, and on the right panel - the hole rotating with the maximum
angular velocity. The result are shown for the orbits of the semimajor axis 
$a=50~{\rm AU}$ (thin), $150~{\rm AU}$ (medium), and $250~{\rm AU}$ (thick
lines). 
}
\end{figure}
\begin{figure}[p]
\vspace{10.0cm}
\includegraphics{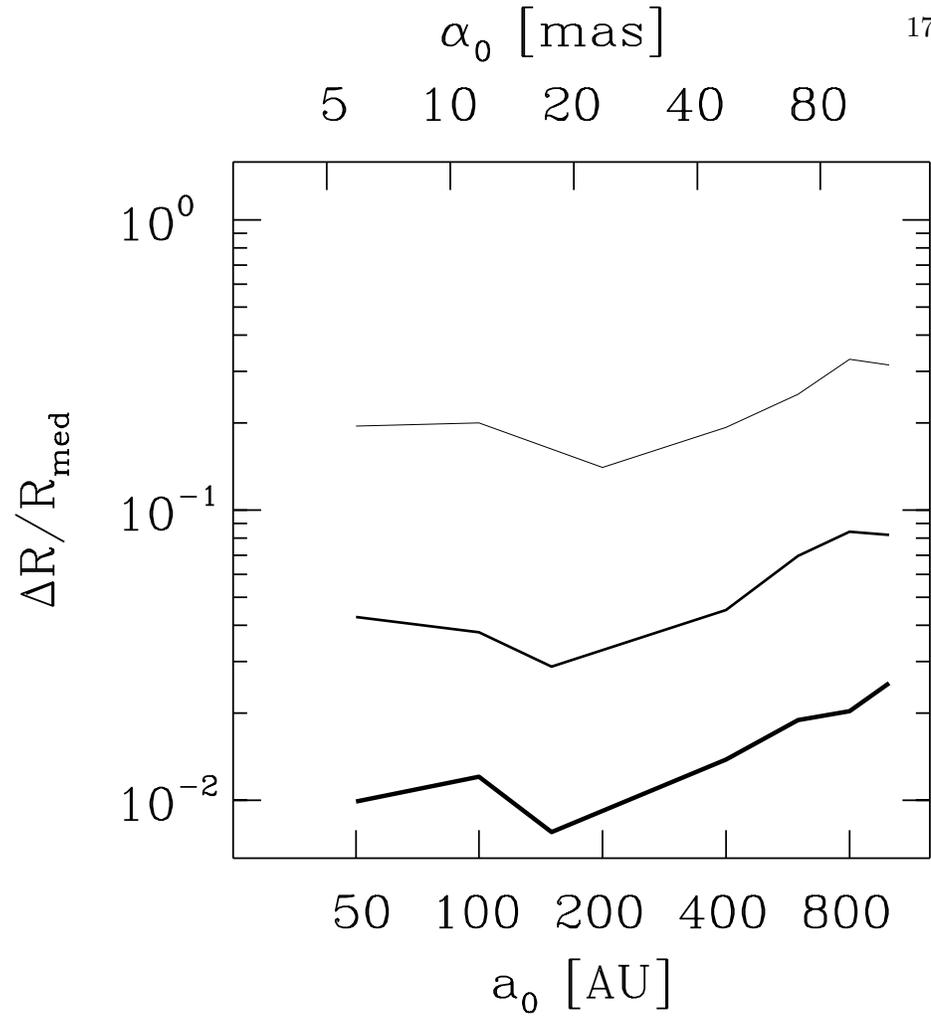}
\caption{The typical error in the distance determination based on the
measurement of periastron motion as a function of the orbit  size.
The results are shown for the astrometric accuracy of $20~{\rm \mu as}$
($0.17~{\rm AU}$, thick line), $80~{\rm \mu as}$ ($0.68~{\rm AU}$, medium), and
$0.4~{\rm mas}$ ($3.4~{\rm AU}$, thin).
}
\end{figure}

\end{document}